# Paradox of Peroxy Defects and Positive Holes in Rocks

# Part II: Outflow of Electric Currents from Stressed Rocks

**John Scoville**[1,2,3], **Jaufray Sornette**[3], and **Friedemann T. Freund**[1,2,4]

[1] GeoCosmo Science and Research Center, Mountain View, CA 94043

[2] San Jose State University, Dept of Physics, San Jose, CA 95192-0106

[3] NASA Ames Research Center, Moffett Field, CA 94035

[4] SETI Institute, Mountain View, CA 94043

Corresponding author Friedemann Freund **friedemann.t.freund@nasa.gov**

**Abstract**

Understanding the electrical properties of rocks is of fundamental interest. We report on currents generated when stresses are applied. Loading the center of gabbro tiles, 30x30x0.9 cm$^3$, across a 5 cm diameter piston, leads to positive currents flowing from the center to the unstressed edges. Changing the constant rate of loading over 5 orders of magnitude from 0.2 kPa/s to 20 MPa/s produces positive currents, which start to flow already at low stress levels, <5 MPa. The currents increase as long as stresses increase. At constant load they flow for hours, days, even weeks and months, slowly decreasing with time. When stresses are removed, they rapidly disappear but can be made to reappear upon reloading. These currents are consistent with the stress-activation of peroxy defects, such as $O_3Si$-OO-$SiO_3$, in the matrix of rock-forming minerals. The peroxy break-up leads to positive holes h$^•$, i.e. electronic states associated with O$^-$ in a matrix of O$^{2-}$, plus electrons, e'. Propagating along the upper edge of the valence band, the h$^•$ are able to flow from stressed to unstressed rock, traveling fast and far by way of a phonon-assisted electron hopping mechanism using energy levels at the upper edge of the valence band. Impacting the tile center leads to h$^•$ pulses, 4-6 ms long, flowing outward at ~100 m/sec at a current equivalent to 1-2x10$^9$ A/km$^3$. Electrons, trapped in the broken peroxy bonds, are also mobile, but only within the stressed volume.

1. Introduction

The electrical properties of rocks are of fundamental interest for many aspects of the physics of the solid Earth [*Fuji-ta et al.*, 2004; *Nover*, 2005]. This includes processes that can be linked to earthquake and pre-earthquake conditions. Surprisingly, wide gaps remain in the knowledge base, in particular with respect to understanding how electric currents are generated in the rocks at those depths, about 35-45 km, where the majority of destructive earthquakes occur and where the temperatures along the geotherm typically increase to about 600°C.

In Part I of this two-part paper we dealt with changes in electrical conductivity upon heating to about 600°C. In this Part II we deal with stress and how the application of uniaxial stresses affects the electrical properties of rocks. The goal is to contribute to our understanding of the processes that take place when tectonic stresses in the Earth's crust wax and wane, changing the electrical properties of rocks and how this may lead to identifiable signals, which the Earth sends out prior to major earthquakes.

At the outset we note that in the seismogenic zone, between 5-7 km to 35-45 km depth, where some 85% of destructive earthquakes occur, the rocks are essentially all igneous or high-grade metamorphic, meaning that they have either crystallized from magmas or experienced elevated temperatures as part of their geological history. Hence, at high temperatures, where thermodynamic equilibrium conditions prevail, minerals in these rocks have dissolved, i.e. incorporated in their crystal structures, small amounts of gas/fluid components forming solid solutions, ss. The gas/fluid component of prime interest is $H_2O$, which enters the host mineral structures in the form of hydroxyl, $O_3Si-OH$. During cooling the solid solutions inevitably drift out of thermodynamic equilibrium and become supersaturated solid solutions, sss. It is in this non-equilibrium state that a long overlooked redox conversion occurs, in the course of which the two hydroxyl oxygens transfer each on electron onto the hydroxyl proton, thereby turning two $H^+$ into $H_2$ and oxidizing the oxygen to the 1– valence state. The two $O^-$ combine to form a peroxy bond:

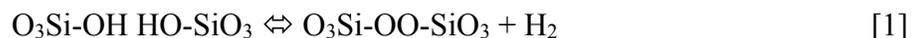
$$O_3Si-OH\ HO-SiO_3 \Leftrightarrow O_3Si-OO-SiO_3 + H_2 \qquad [1]$$

Part I was dedicated to the break-up of the peroxy bond upon heating. Part II will look at the same break-up reaction using stress.

Under Earth's surface conditions, crustal rocks are fairly good electrical insulators. A common way to address electrical properties of rocks in laboratory experiments is to measure their electrical conductivity. This is generally done by placing a rock sample, often a disk or a cylinder with flat faces as depicted in **Figure 1a**, between two electrodes. A voltage is then applied to produce an alternating current (ac) in the range of kHz to MHz or a direct current (dc). An ac voltage causes charges or dipoles to oscillate forth and back within the sample. A dc voltage will drive mobile charge carriers through the sample. A guard electrode may be added to catch currents flowing along the sample surface, allowing to separately monitor the currents flowing along the surface and through the bulk. Stresses may be applied, either uniaxially or triaxially [*Heikamp and Nover*, 2001].

The term "stress" is often used in somewhat imprecise ways. The dimension of stress is that of pressure in units of Pascals (Pa), given as Newtons per $m^2$. Solids can support both normal (compressional) and shear stresses where normal stresses are perpendicular to the surface and shear stresses parallel to it. Loading a cylinder uniaxially creates mostly normal stresses but there is always also a shear stress component. In the context of this Part II the shear stress components are most important.

Applying mechanical stresses to igneous or high-grade metamorphic rocks has long been recognized to lead to an increase in electrical conductivity, particularly during shock-loading [*Gorshkov et al.*, 2001]. Several explanations for the increased conductivity have been offered in the literature such as improved grain-grain contacts [*Glover*, 1996; *Nover*, 2005] or thin graphite layers forming spontaneously on internal crack surfaces [*Nover et al.*, 2005].

Figure 1b illustrates an experimental set-up of the type described in the literature to measure pressure-stimulated voltages or currents, which the voltages generate [*Aydin et al.*, 2009; *Johnston,* 1997; *Kyriazis et al.,* 2009; *Triantis et al.*, 2006; *Tullis,* 2002; *Vallianatos and Triantis,* 2008; *Triantis and Vallianatos*, 2012]. Generally the rock cylinders or disks are loaded over their entire cross sections [*Brace,*1965, 1968; *Zhu,*2001]. Due to unavoidable inhomogeneities in the set-up or in the local materials properties, this will lead to nonuniform distributions of normal and shear stresses across of the rock samples. They manifest themselves in erratic potential differences between different parts of the stressed rock cylinders, making it difficult, if not impossible, to use those potential differences to derive information about the sign and the flow of charge carriers.

On a dynamically active planet like Earth, stress gradients are the most characteristic collateral of pre-earthquake conditions. Stress gradients act over a wide range of distances. They wax and wane on many time scales [*Zoback et al.,* 1987; *Sornette et al.,* 1994; *Seeber and Armbruster,* 2000; *Anderson and Ji,* 2003; *Kagan et al.,* 2005]. If we suspect that stresses can cause electric currents, it is advisable to stay away from applying external voltages to the rock sample. Instead we should set up laboratory experiments in such a way that stress gradients become an integral part of the design and that stress-generated currents (and voltages) can be measured. After all, while Nature is known to produce ample stress gradients across rocks in the Earth's crust, Nature does not apply specific voltages, in particular not ac voltages in the kHz to MHz range such as often used in conventional electrical conductivity measurements.

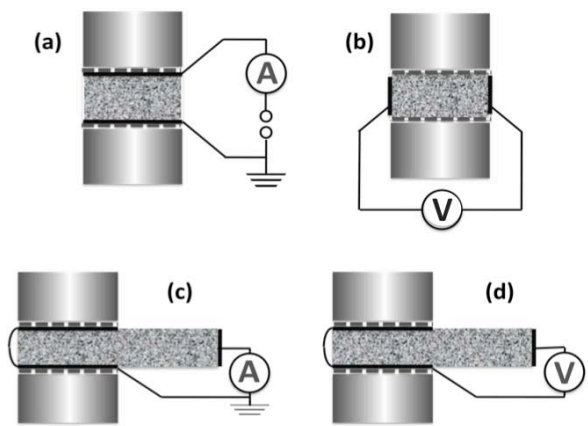

*Figure 1a*- *Exemplary set-up for measuring conductivity of a rock subjected to uniaxial load, where the current is driven through the sample by an externally applied dc or ac field.*

*Figure 1b*- *Exemplary set-up for measuring a pressure-stimulated voltage in a quasi-uniformly loaded cylindrical rock sample.*

*Figure 1c/d*- *Set-up for measuring currents and electrical potentials along a stress gradient without externally applied electric potentials.*

*Dashed lines: insulator sheets*
*Solid lines: metal contacts*

Hence, instead of applying an external electric field to drive existing charge carriers through the rock sample as depicted in Figure 1a or using a quasi-uniformly loaded cylindrical sample to cause stress-stimulated charges to build up hard-to-interpret voltages as depicted in Figure 1b, we subject one end of a rock sample to uniaxial load as shown in Figure 1c/d, leaving the other, unloaded end stress-free. Placing a voltmeter or an ammeter into the electric circuit we capture self-generated voltages and currents, activated by stress, between the stressed and unstressed ends of the rock. Experiments as sketched in Figure 1c/d closely imitate real situations in the field, because stress gradients are part of any system in the Earth's crust and stress gradients are of interest when focusing on pre-earthquake processes. We thus use stress gradients as the thermodynamic driving force causing mobile charge carriers to flow.

An experimental set-up as shown in Figures 1c/d was used earlier to provide information about charge carriers that become stress-activated, and out flow of the stressed subvolume leading to the concept that rocks, when stressed, behave like a battery [*Freund et al.*, 2006; *Takeuchi et al.*, 2006]. Here we focus on the rate at which stresses are applied and the magnitude of the currents.

Prior to the formulation of the battery concept, impact experiments had been conducted where cylinders of various igneous rocks or blocks of granite were hit with projectiles over the velocity range from about 100 m sec$^{-1}$ to 1.5 km sec$^{-1}$ with some experiments up to 4.5 and 5.6 km sec$^{-1}$ [*Freund*, 2002]. The low velocity impacts suggested that positive charges were created by the burst of stress delivered to the point of impact, able spread without externally applied voltage along the cylinder axis at speeds of 200±100 m sec$^{-1}$ [*Freund*, 2002]. In the case of higher velocity impacts the seismic waves propagating from the impact point led to an instantaneous activation of positive charges throughout the rock volume. Due to Coulomb repulsion these charge carriers began to redistribute inside the rock, spreading to the rock surface at their characteristic speed, leading to a positive surface charge, which in turn led to surface/subsurface electric fields and to follow-on reactions.

Geological stress gradients develop where tectonic plates are shifting relative to each other. The gradients usually evolve slowly with time. In pre-earthquake situations stress gradients will evolve increasingly fast as the time of catastrophic failure approaches. In fact, in any rock volume deep in the Earth's crust, which is destined to become the hypocenter of an earthquake, failure under the overload of 10-30 km of rocks will differ significantly from the breaking of rocks in laboratory experiments where rock cylinders, usually unconfined, are loaded. Because of the lack of confinement, the rock cylinders can bulge outward, reduce the overall stress by increasing the volume, and initiate failure at the surface by tensile stress.

By contrast, rock volumes deep in the Earth's crust can hardly increase in volume. The build-up of *in situ* stresses is primarily controlled by the repulsive interactions between atoms and ions that are pushed closer together by the external forces. These repulsive interactions scale with the inverse of $r^9$ to $r^{12}$, where r are the interatomic distances. This very steep increase causes the *in situ* stresses to evolve over increasingly shorter timescales until failure occurs by shear [*Freund and Sornette*, 2007]. For this reason, in this Part II, we report on rock stressing experiments for which the rate of loading was increased from slow to very fast, spanning 8 orders of magnitude.

## 2. Experimental Method

### 2.1 Samples

Gabbros are the depth equivalent of basalts, the most common effusive igneous rock formed from magmas that rise from upper mantle depths [*Green and Ringwood*, 1967]. We conducted our experiments with rock tiles, 30 x 30 x 0.9 cm$^3$, stressed in the center. The tiles consisted of fine-grained, black gabbro from Shanxi, China, commercially available under the trade name "Absolute Black". One side was polished, the other side and the edges were saw-cut rough. The modal composition of this gabbro[*] is approximately 40% plagioclase (about 80% albite and 20% anorthite), 30% clinopyroxene (augite) with minor alteration to amphibole and/or chlorite, and 25% opaques, probably magnetite, plus rims of a reddish mineral, possibly iddingsite. Because they are commercially available, we were able to obtain a large batch of visually homogeneous tiles from the same location.

Identical experiments were conducted with regular plate glass, Ca-Na-aluminosilicate, of the same thickness, 0.9 cm, cut to the same dimensions, 30 cm x 30 cm, and fitted with a pair of 5 cm diameter Cu electrodes in the center and the same Cu electrode continuously along the edges.

**2.2 Hydraulic Press Experiments**

We conducted tests at slow to moderately fast, constant loading rates in an IPC Universal Testing Machine, UTM 100, capable of delivering 100 kN load, controlled at the ±1 N level[†]. To reduce the electric noise we converted the UTM 100 environmental chamber into a Faraday cage by covering the glass window door with grounded aluminum (Al) foil. All leads to the rock samples were either BNC cables or Cu leads shielded with grounded Al foil.

The tiles were centrally loaded between two steel pistons, 3.9 or 5.0 cm diameter (Figure 2a/b). The pistons were electrically insulated from the press by 0.5 mm thick polyethylene sheets. Copper contacts with graphite-based adhesive (3M Corporation), were applied to the rock surface underneath both pistons. Cu stripes, 0.67 cm wide, were applied along the edges of the tiles. The outflow currents were measured between the central stressed volume and the edges of the tiles. For recording potential differences between the stressed rock volume and the surface of the tiles we used a large-area capacitive sensor consisting of Cu tape applied to the backside of a 0.5 mm thick sheet of polyethylene lying loosely on the polished side of the tile.

---

[*] Courtesy of the late Paul Lohman, Geodynamics Laboratory, NASA Goddard Space Flight Center
[†] Courtesy of Charles Schwartz, Dept. of Civil Engineering, University of Maryland

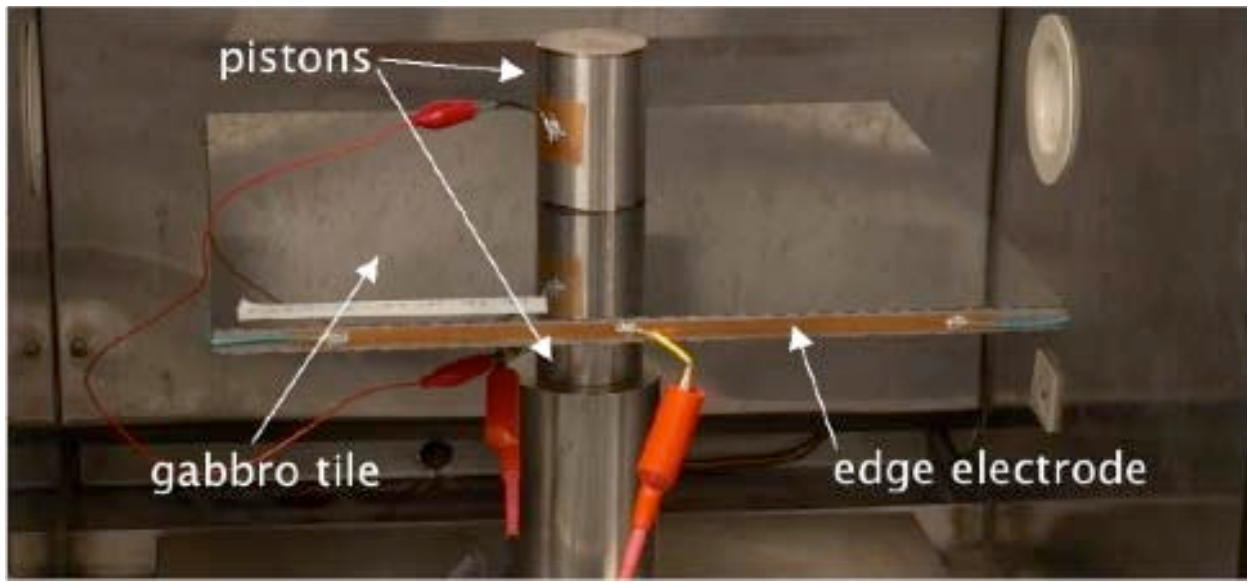

*Figure 2* - *Photograph of a gabbro tile inside the Faraday cage set up to be loaded through two pistons in the center.*

For current measurements, the HI (high) side of a Keithley 617 electrometer was connected to the edge Cu electrode, while the LO (low) side was connected to the center contact and to ground (Figure 2b). For voltage measurements, the same Keithley 617 was used in voltage mode. For data acquisition we used LabVIEW (National Instruments Corp.). As a rule we allowed the rock tiles to anneal without load for several hours or overnight before loading or reloading.

We loaded the tiles from 0 MPa to pre-set maximum stress levels, using fast loading at a pre-set level or slow loading at constant rates, varying over 5 orders of magnitude, from 0.2 kPa/sec to over 60 MPa/sec. Typical maximum stress levels ranged from 48 to 63 MPa, corresponding to less than 1/3 the load necessary to cause unconfined gabbro or the plate glass to fail.

### 2.3 Free-Fall Acceleration Experiments

A Monterey Research Laboratory, Inc. Model Impac 66 drop tower was used to load the gabbro tiles more rapidly, within 0.5-2 ms by dropping a 90.7 kg aluminum weight from various heights onto a 2-ton spring-mounted steel body. We dropped the weight onto a 5 cm diameter steel piston placed on a 5 cm diameter Cu electrode at the tile center as shown in Figure 2b, cushioning the impact with two to three felt pads. The rebound of the drop mass was halted by air brakes. The drop height was varied from 5 cm to about 25 cm, resulting in accelerations from 250 g to 1100 g (2,450-10,780 m/s$^2$) over 0.5-2 ms, corresponding to a forces of 220 to 980 kN, as measured with an accelerometer installed on the drop mass. The force during impact was distributed over an area of ~20 cm$^2$ leading to peak pressures ranging from 100 to 500 MPa.

Figure 3 shows a tile being readied to a drop experiment. The Cu contact along the edges can be seen. Beneath the 5 cm diameter piston is a 5 cm diameter Cu contact on the topside of the tile and an identical Cu contact on the underside. Kapton sheets, 50 μm thick, were used to electrically insulate the tile from its support and from the drop mass. Kapton also served to insulate thin Cu stripes between two Kapton sheets to contact the 5 cm diameter electrodes at the center of the tile, on the top and the bottom face.

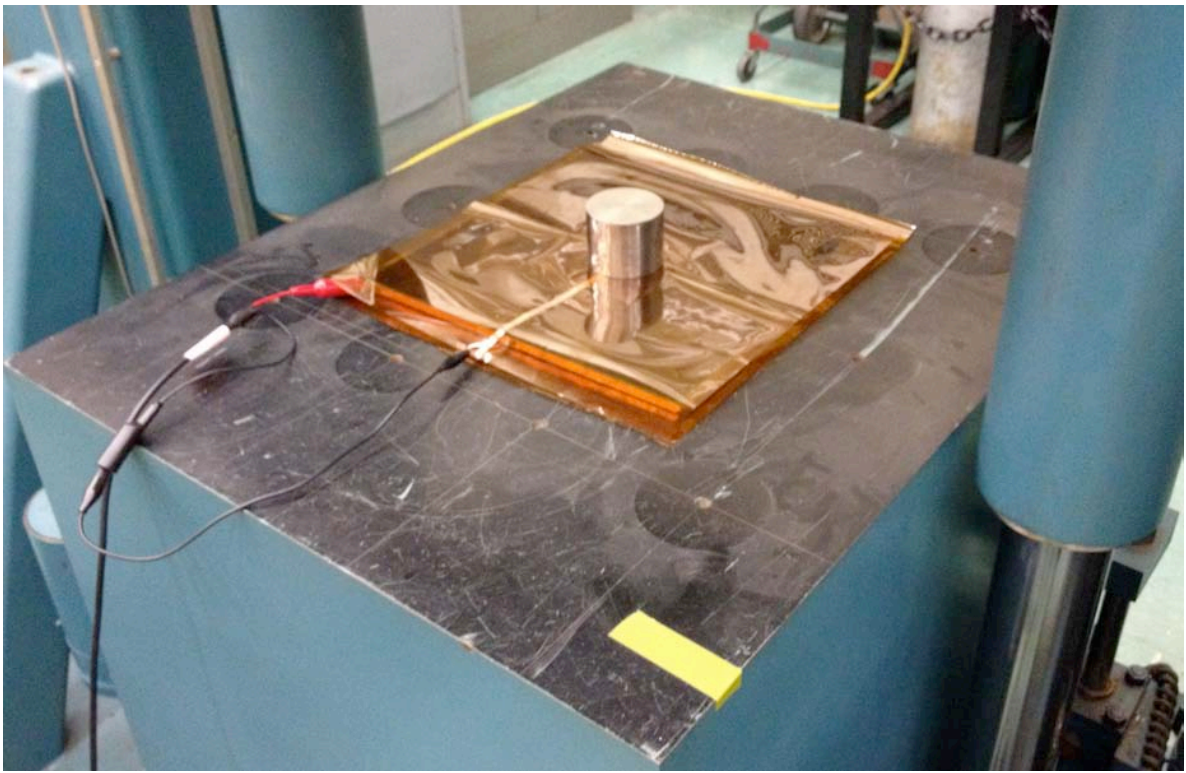

*Figure 3- Experiment configuration for a drop-tower experiment with impact to take place on a 5 cm diameter stainless steel piston on the center of the rock tile to measure the flow of charge carriers from the stressed rock volume underneath the piston through unstressed rock to the edge. The red connector goes to the edge contact, while the black connector goes to the tile center. Kapton sheets are used as insulators.*

Kinetically, the drop experiments differed from the runs with the hydraulic press in as much as the rate of stress increase during the impacts was essentially parabolic. After each drop test the tiles were inspected for cracks, and the accelerometer output was inspected for signs of cracking as evidenced by jagged leading edges on what would otherwise show as smooth, approximately parabolic acceleration curves. High impact pressures in combination with minimal padding often caused the tiles to crack, resulting in mechanical and radiative energy dissipation in addition to noisy data. If a tile was found to have fractured, its data were excluded from further analysis. By reducing the drop height we obtained usable data with and without padding. After identifying the right conditions under which the tile survived the impacts, we subjected this tile to several consecutive impacts, separated by only a few minutes between drops.

Currents were recorded by connecting the black terminal of a Stanford Research Systems SR570 preamplifier to the 5 cm diameter Cu contacts at the center of the tile, on both sides, and its red terminal to the Cu tape around the edges of the tile. The preamplifier currents were translated into voltages using the SR570 feedback ammeter. Typical settings from 5 to 100 µA/V. Using a Picoscope SB3206A the feedback ammeter output was transferred to a computer oscilloscope. The voltages were scaled by the appropriate preamplifier settings to measure the currents.

## 3. Results

### 3.1 Hydraulic Press Experiments

First we report on an experiment with the plate glass tile, letting it sit for several hours at room temperature to allow any charges that might have accumulated to dissipate. As shown in Figure 4a there was a small current in the 30 pA range, slowly decreasing with time. We loaded the glass tile from 0 to 50 MPa within 1 sec and kept the load constant for over 2 hrs. During loading we observed no change in the current.

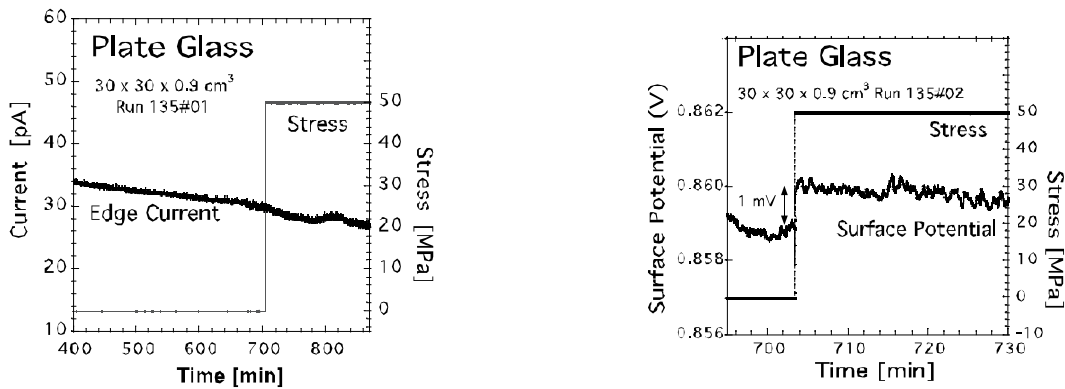

*Figure 4a/b: Current and surface potential changes of a plate glass tile, cut to the same dimensions as the rock tiles, during rapid loading from 0 to 50 MPa.*

We repeated the experiment, after letting the glass tile sit overnight, this time measuring the voltage. As shown in Figure 4 b, just before loading, the surface potential was close to +859 mV.

During loading it shifted by 1 mV to +860 mV.

**(a)** 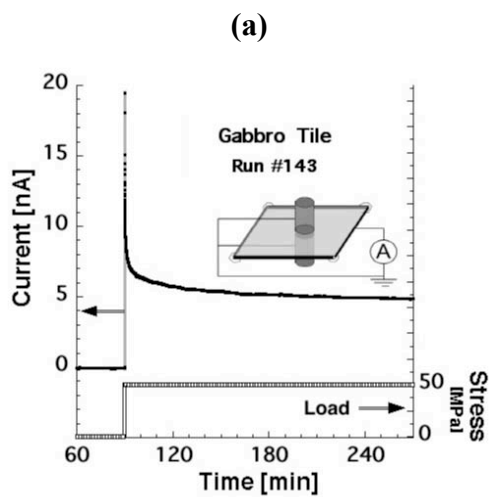

**(b)** 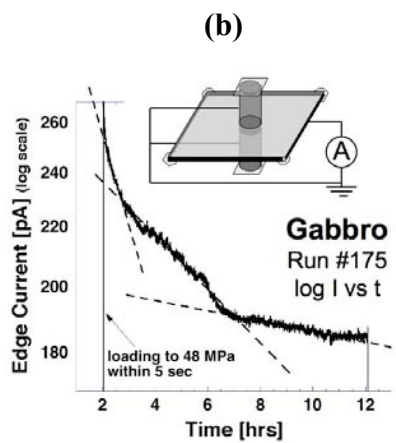

**(c)** 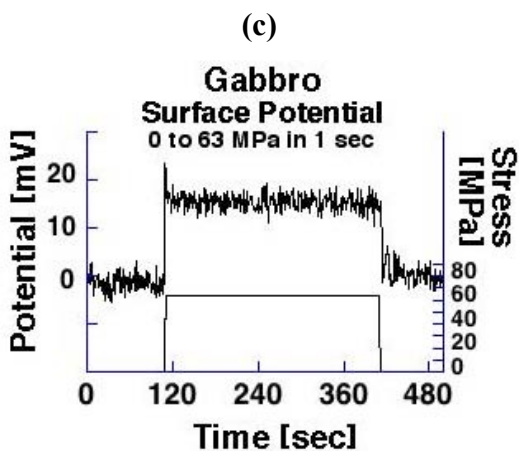

*Figure 5a*- Load and current versus time during loading of a rock tile to 48 MPa within 1 sec, followed by 3 hrs at constant load.

*Figure 5b*- Decay of the positive outflow current over 10 hrs at a constant load of 48 MPa, plotted on a semilogarithmic scale.

*Figure 5c*- Surface potential versus time during loading of rock tile within 1 sec to 58 MPa, holding the load constant from 50 sec and releasing it at the same rate.

Figure 5a/b show the dramatically different current when subjecting a gabbro tile to the same procedure. Before applying the load the background current between center and edges was on the order of 10 pA. We loaded the tile within 1 sec from 0 to 48 MPa and then kept it constant for about 100 min as shown in Figure 5a. Upon loading an instant current was recorded, increasing rapidly almost 10,000-fold, to about 20 nA. At constant load this current decayed, rapidly at first to about 6-7 nA, then more slowly to about 5 nA.

Repeating the experiment but keeping the time at constant load 10 hours and plotting the current on a logarithmic scale versus linear time t, produces a decay curve with three straight sections with different slopes as shown in Figure 5b. These three sections suggest that the stress-activated current flowing out of the central rock volume is carried by charge carriers that anneal with different halftimes.

Following convention, the positive direction of a current is when electrons flow in the opposite direction [*Berlin and Getz*, 1988]. This means that the positive direction of a current is the direction of the flow of holes. The centrally loaded gabbro tiles consistently produced positive currents from the stressed center to the edges, indicating outflow of holes.

Figure 6a shows the current during a full cycle of loading and unloading a gabbro tile at a moderate constant rate, 6 MPa/min to 48 MPa, keeping the load constant for 30 min, followed by ramping down at 6 MPa/min. Figure 6b shows the voltage during a similar cycle at different run parameters (rate, maximum load, dwell time). The results of both experiments are very similar with respect to the way the current and the voltage evolve.

The outflow current from the centrally stress rock volume is positive, and the potential difference between center and the surface of the tile is such that the surface becomes positive relative to the stressed center volume. Several additional features are noteworthy for the current:

(i) The current before loading is negligibly small, a few hundred fA to a few pA, often slightly different from tile to tile despite apparent homogeneity of the gabbro matrix;

(ii) Upon loading the center of the tile, the current starts with a small excursion to negative values but then quickly turns positive, raising fast already at low load levels, reaching currents typically 100x stronger than the baseline current;

(iii) The current passes through maxima and minima between 5-10 MPa while being loaded at the constant rate and saturate above 20 MPa;

(iv) At constant load, 48 MPa, the current decreases slowly and steadily;

(v) Upon unloading, similar current maxima and minima develop as during the loading phase with the main decrease in current only after the load has dropped to about the same value as observed during the rapid increase of the current during loading;

(vi) After the load has returned to zero, a small residual current often flows for hours before returning to the initial value in the few hundred fA to low pA range.

In the case of the voltage measurement depicted in Figure 6b the gabbro tile, stressed at 12.6 MPa/min to 63 MPa and held for 5 min before unloading, develops a brief negative voltages at the very beginning and the very end, transitioning to a positive potential difference between the stressed center of the tile and the unstressed part. The voltage reaches its highest value around +10 mV soon after the beginning of loading, then drops to around +5 mV, increasing slowly as long as the stress increases. Once the stress level becomes constant, the potential difference decreases slightly, similar to the way the current decreases under constant load. The voltage reaches a high value as second time during unloading, when the load is already almost completely removed.

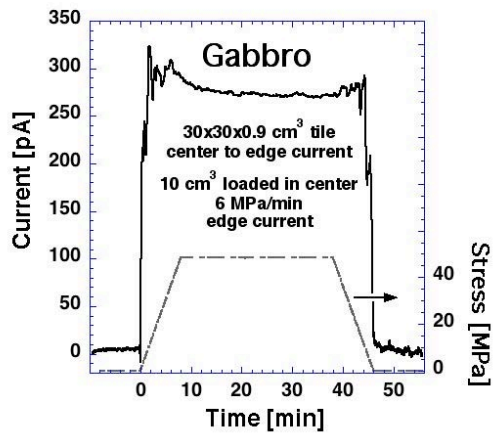 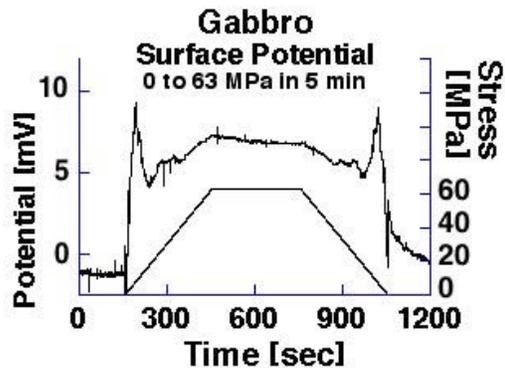

*Figure 6a- Outflow current from a centrally stressed gabbro tile during a loading-unloading cycle to 48 MPa at a rate of 6 MPa/min.*

*Figure 6b- Potential difference between the center of a gabbro tile and the rock surface during loading-unloading at 12.6 MPa/min.*

The similarity between voltage and current during the loading cycles indicates that the voltage, marking the self-generated potential difference between the stressed center to the unstressed edges, drives the current outflow.

The potential, here measured as voltage, is self-generated. Hence, the current is self-generated, flowing in response to a self-generated potential difference, not in response to an externally applied electric field. The fact that the highest currents and voltages are recorded at low stress levels, less than 5-10 MPa, calls into question explanations offered in the literature for pressure-stimulated currents and voltages, in particular the suggestion that they may be the result of improved grain-grain contacts [*Glover*, 1996; *Nover*, 2005] or of internal crack formation with carbon or graphite films forming spontaneously on the fracture surfaces [*Nover et al.*, 2005].

If a gabbro tile is repeatedly loaded and unloaded, it produces nearly fully reproducible current profiles. An example is given in Figure 7, where a gabbro tile was subjected over a period of 10½ hrs to 22 identical loading-unloading cycles from 0 to 48 MPa. Each cycle generated nearly identical outflow currents, consistently positive with maxima at low stress levels and at the end of each unloading. At constant 48 MPa stress the current reaches 240 pA, decreasing slightly, less than 5%, after 22 loading-unloading cycles.

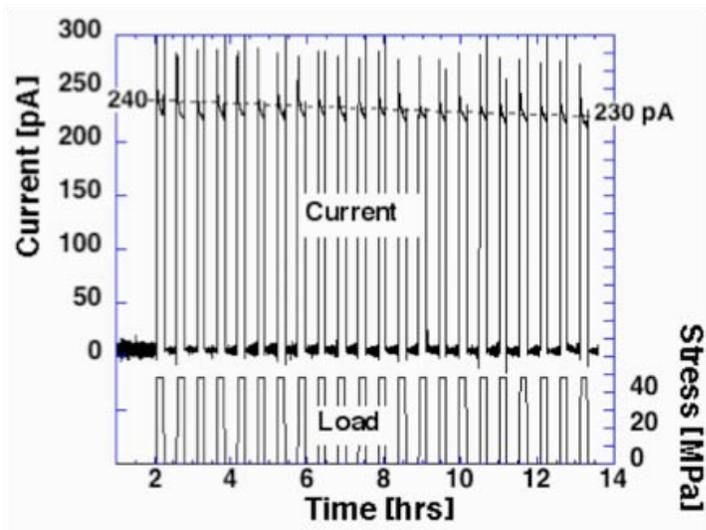

Figure 7- Sequence of 22 loading-unloading cycles from 0 MPa to 48 MPa of a centrally loaded gabbro tile.

It is remarkable that, once an elevated stress level has been achieved, there is a near-constant current flowing out of the stressed rock volume, decreasing only slowly with time. By contrast, when the load is removed, the outflow current drops precipitously when the load level has reached values below 5-10 <MPa. This observation is consistent with the concept that the rock represents a battery that is being "turned on" by applying a load and "turned off" by removing the load [*Freund et al.*, 2006; *Takeuchi et al.*, 2006]. If the resistance in the outer circuit is low, the amount of current that can flow out of the stressed rock volume is controlled by the self-generated voltage and the internal resistance of the rock.

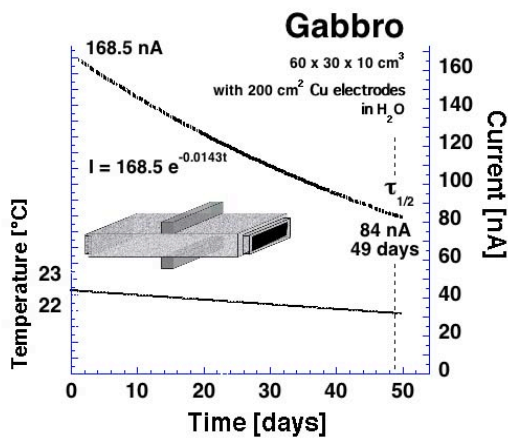

Figure 8- Applying a constant load to a large block of gabbro and keeping it for 50 days leads to a continuous outflow current, which decays with a half-life on the order of 49 days.

This is illustrated by Figure 8, derived from a different measurement where a large block of gabbro, 30 x 60 x 10 cm$^3$, was loaded along its center line to 28 MPa, using a pair of steel bars as pistons with a 30 x 5 cm$^2$ footprint. Applying the load produced an outflow current on the order of 170 nA. Keeping the load constant for 50 days, led to a steady decrease of the current to about one half its starting value after 49 days. The number of charge carriers flowing out of the 150 cm³ stressed volume over this time period is about $3 \times 10^{18}$, equivalent to ~0.5 Coulomb.

To illustrate in more detail the onset of the self-generated currents flowing out of a 30x30x0.9 cm³ gabbro tile during loading we magnify in Figures 9a-c the initial sections at three loading rates, 6 MPa/min, 60 MPa/min, and 600 MPa/min, marking arbitrarily, to aid the eye, the stress levels 5 MPa, 10 MPa, and 20 MPa.

**(a)**

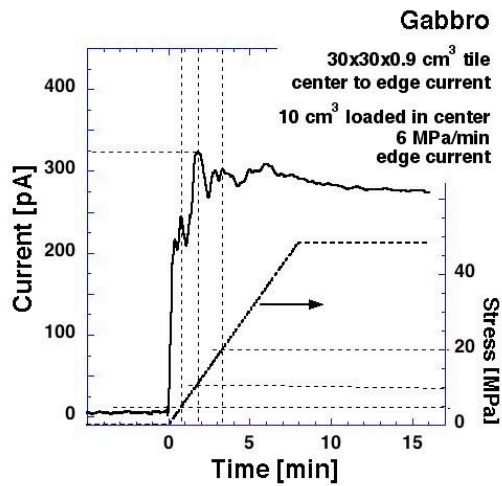

**(b)**

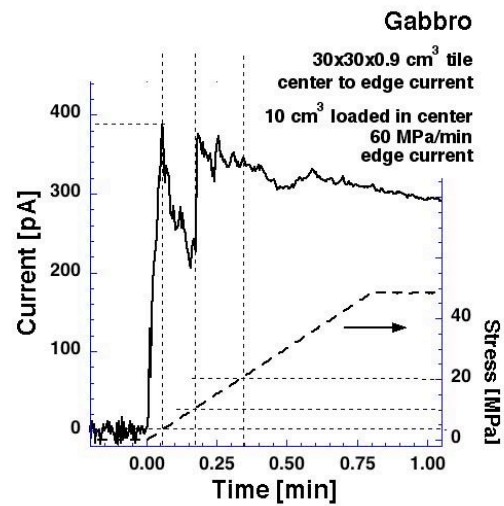

**(c)**

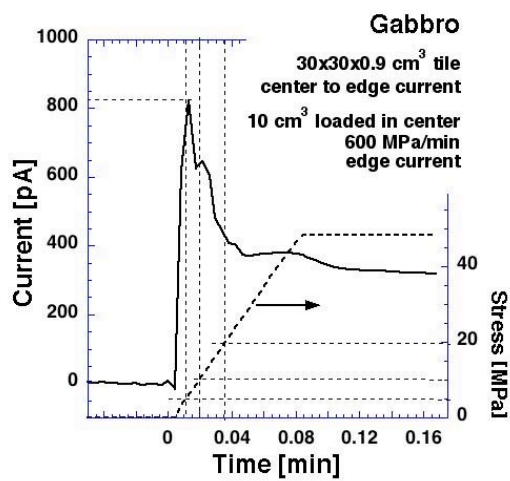

*Figure 9: A characteristic feature of the stress-activation of charge carriers in rocks are the rapid increase of the outflow current and the intense current fluctuations during the early phase when the gabbro tile is loaded at the constant rates:*

*(a) at 6 MPa/min,*

*(b) at 60 MPa/min,*

*(C) AT 600 MPA/MIN.*

One striking feature of Figures 9a-c are current maxima already at modest stress levels, ~5 MPa at loading rates 60 and 600 MPa/min. At the slower rate, 6 MPa/min, a first current maximum is reached at 5 MPa and a second, higher one at 10 MPa. Another striking feature is that the initial current maxima are higher than the currents flowing under steady load and they increase significantly in magnitude with the fast the rock tile is loaded. In summary, the characteristic features of the outflow of charges from a stressed subvolume are

(i) Early onset of current outflow already at very low loads,
(ii) transient current maxima during the initial phase of the loading,
(iii) increase of the current maxima with increasing rate of loading,
(iv) current fluctuations after the initial maxima,
(v) slow decrease of the current under constant load.

The fact that the currents increase from 325 pA at 6 MPa/min to 390 pA at 60 MPa/min and to 820 pA at 600 MPa/min, suggests a stress-activation of charge carriers with different lifetimes during loading. Some charge carriers appear to be short-lived, deactivating within less than 1 sec, others are long-lived with longer lifetimes. As the example presented in Figure 8 demonstrates, the lifetimes under constant load can be on the order of days to weeks, even months.

This characteristic behavior was further confirmed by setting up an experiment with a new gabbro tile and stressing it repeatedly, first slowly, then faster, covering 5 orders of magnitude of loading rates, from 0.24 kPa/sec to 19 MPa/sec. Upon reaching a time limit or 58 MPa, the tile was unloaded and allowed to anneal for 2-4 hrs or longer before the next cycle.

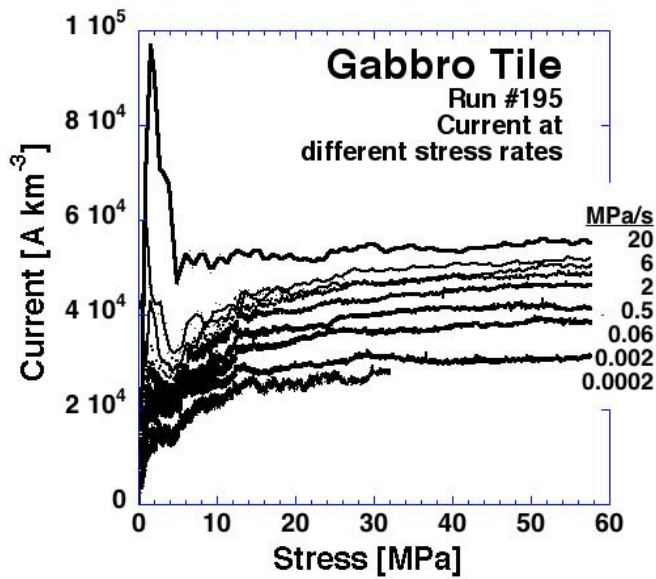

*Figure 10- Outflow currents from a centrally loaded tile at different loading rates spanning 5 orders of magnitude*

Figure 10 shows a set of cycles recorded at increasing constant loading rates up to 58 MPa, which corresponds to about ¼ the stress needed to cause the gabbro to fail. Even though the rock was loaded at constant rates, the outflow currents are not steady. They tend to fluctuate, in particular during the very early loading stage. At slow rates the currents cluster below 5 MPa. They evolve into a sharp peak of high outflow currents low MPa values for the fastest loading rates. In addition Figure 10 shows that (i) the currents increase very fast the faster the loading rate and (ii) the current do not level out but instead continue to increase as long as the loading continues.

Figure 11 shows the same current vs load curves, autoscaled to emphasize the fluctuations at low loads and the evolution of a sharp peak around 5 MPa at fast loading.

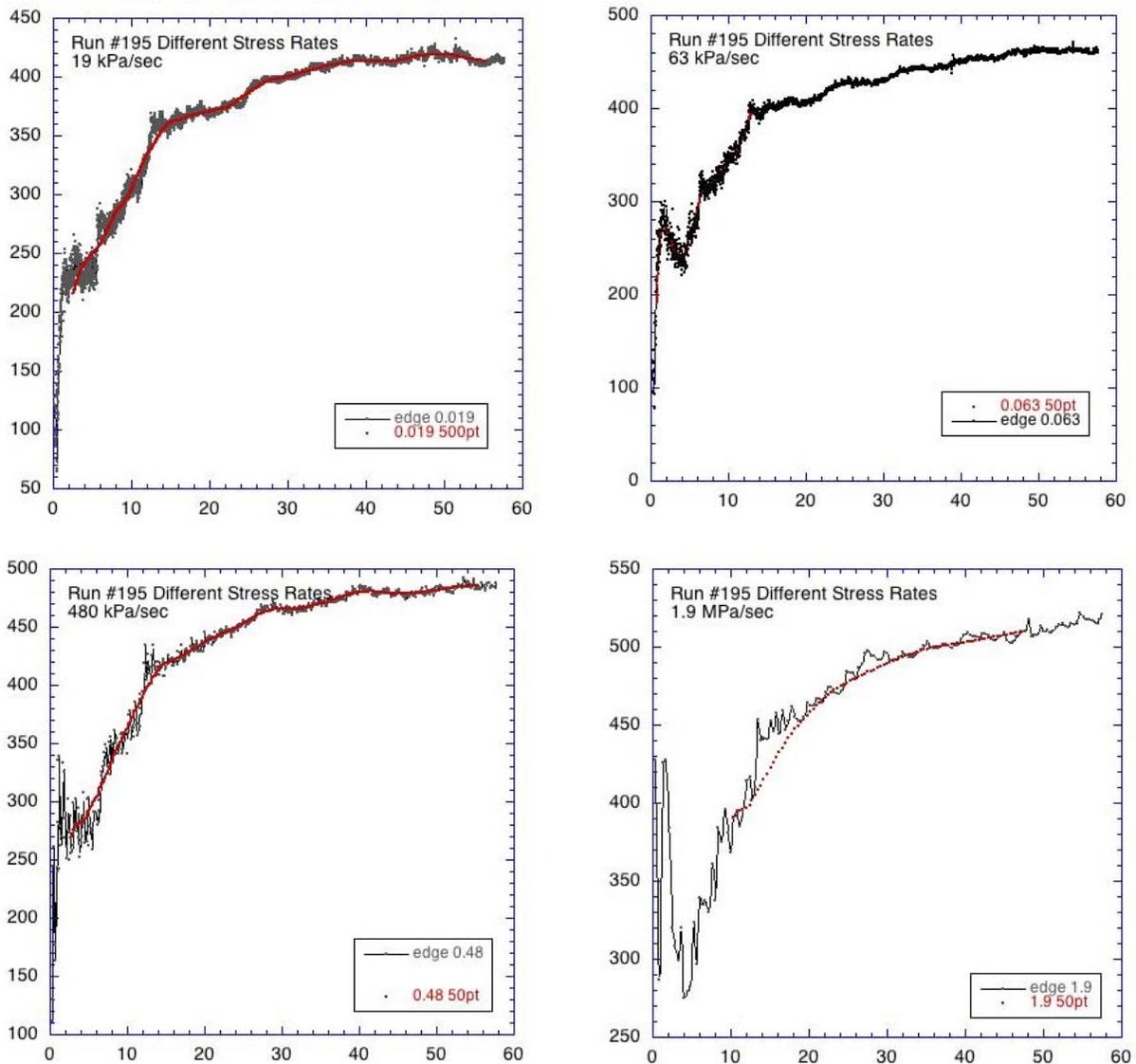

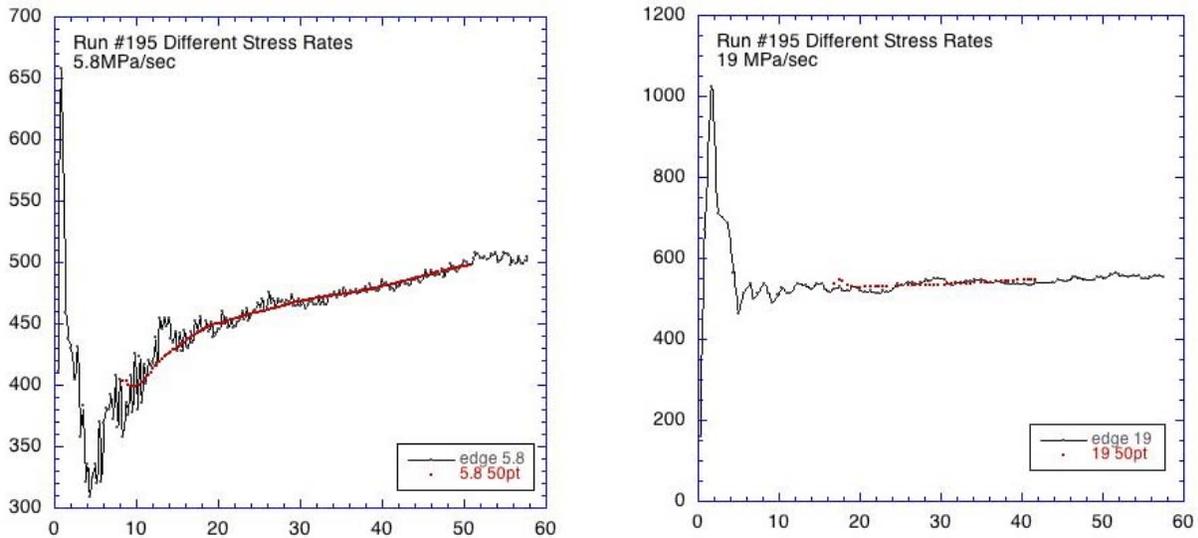

*Figure 11- Outflow currents from a centrally loaded gabbro tile, 10 cm³ stressed rock volume, at different stress rates covering nearly 5 orders of magnitude, up to a maximum load of 58 MPa.*

Figure 12 plots the outflow current per km³ versus log(time). The currents increase with increasing rate, approaching $10^5$ A km³ at the fastest rate. The increase is faster than exponential.

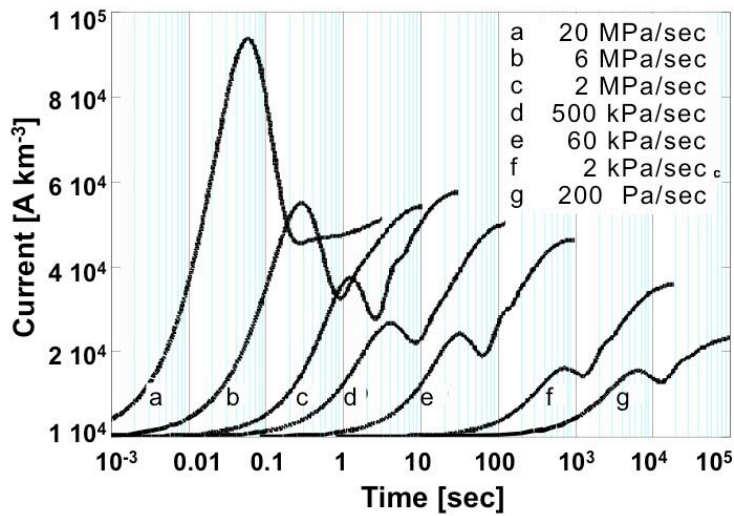

*Figure 12- Outflow currents from a centrally loaded tile at different loading rates spanning 5 orders of magnitude.*

Figure 13 plots the peak currents versus the ramp. The data suggest that, if the stress rate can be further increased, even larger outflow currents might be expected.

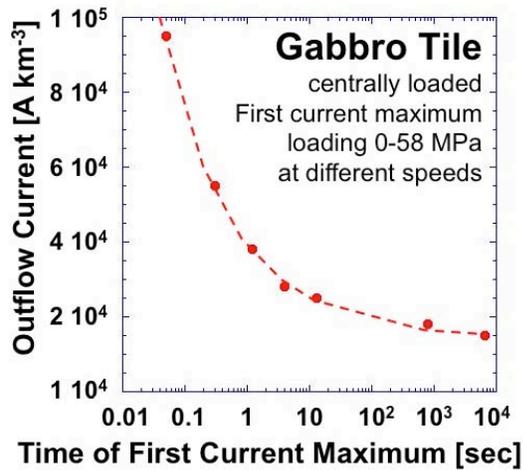

*Figure 13: Plot of the peak current (in A km$^{-3}$) versus the speed with which the load is applied. Curve fitting suggest that, if the ramp can be made even shorter, outflow currents from the centrally loaded gabbro tile equivalent to outflow currents much larger than 10$^5$ A km$^{-3}$ might be achievable.*

**3.2 Drop Tower Experiments**

During the hydraulic press experiments the rate of stress increase was always linear. During the impact experiments the rate was approximately parabolic. Many tiles broke under impact or cracked. By reducing the height from which the 90.7 kg mass was dropped to between 15 and 25 cm and using felt pads to cushion the impact, spreading the time of compaction to about 1 msec, we achieved conditions that did not lead to fracture. Once these conditions were established and a tile survived, we impacted the same tile several times with only a few minutes between drops.

Consistent with the results of the hydraulic press experiments the outflow currents resulting from dropping the 90.7 kg mass on the 5 cm diameter steel piston in the center of the tile were always positive. This confirms that positive holes are the only charge carriers, which can flow from the stressed subvolume at the center of the tile to the outer unstressed region. Electrons arc mobile only within the stressed rock volume [*Freund*, 2013]. After each positive outflow pulse, a weak negative current was observed lasting for about 25-30 msec. This suggests that, after each outflow pulse, some of the positive charge carriers are being pulled back toward the center of the tile. The first drop had a minimal 'reflux' as shown in **Figure 14a**.

**(a)**

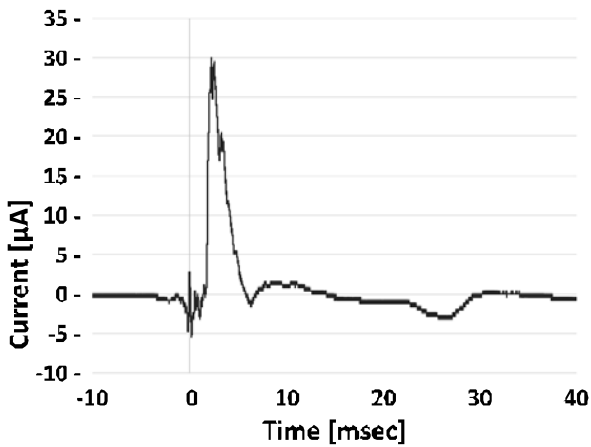

*Figures 14a-e- Currents resulting from centrally loading a tile by dropping a 90.7 kg weight on a piston in the center of the tile. The current pulse was always positive. A weak negative current was observed after the impact over a longer period of time. The first drop on a 'fresh' tile, a, had a minimal 'reflux' of negative current. Subsequent drops, b-e, produced more negative current, but the net currents always remained positive.*

**(b)**

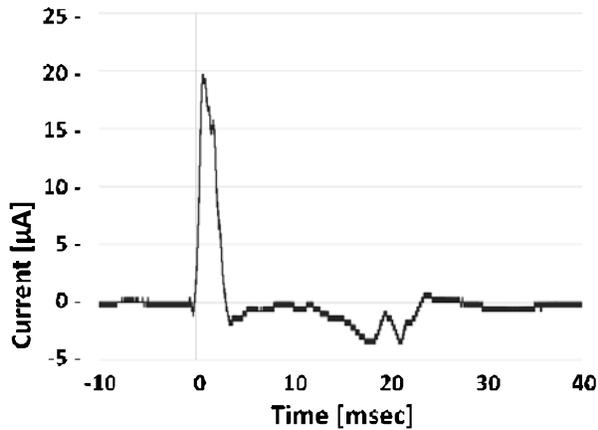

**(c)**

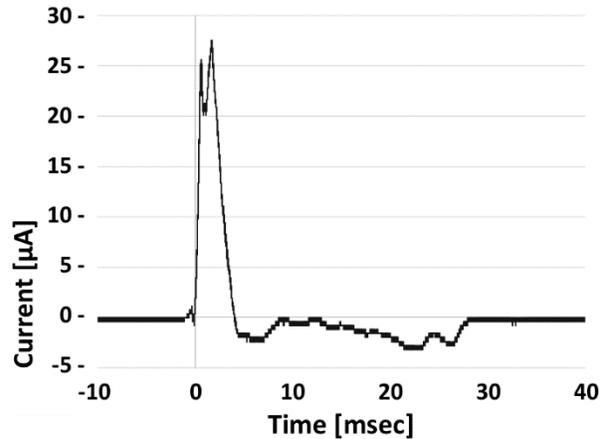

**(d)**

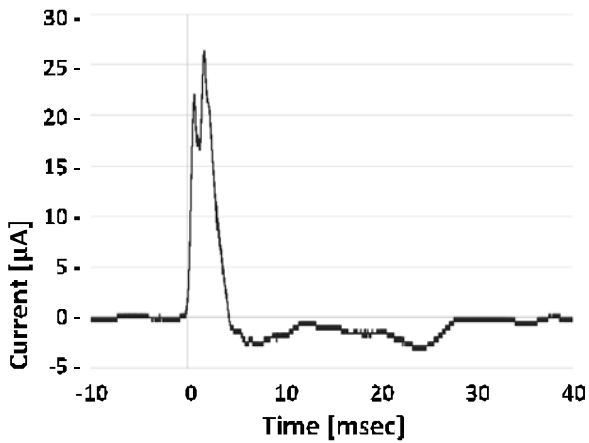

**(e)**

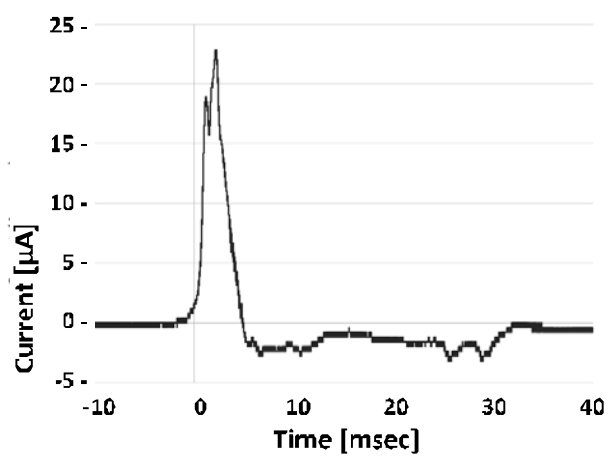

Subsequent drops on the same tile, Figure 14b-e, produced – in the integral – slightly stronger negative currents, but the net current always remained positive as illustrated in Figure 15.

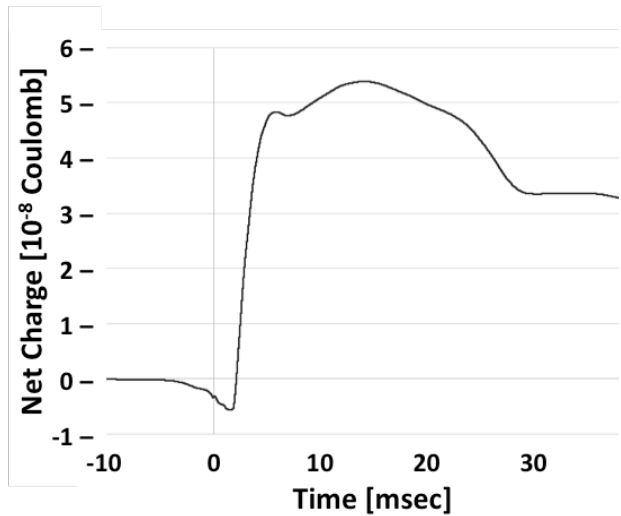

*Figure 15- Net flux of charges during the first drop, resulting from integrating the curve in Figure 14a. Most of the positive charge, which flowed out during the initial pulse, persists at the end.*

Linearly extrapolating the maximum positive outflow current pulse from ~20 cm$^3$ to the current that would flow out of 1 km$^3$ of rock gives 1-2 x 10$^9$ A. This is consistent with the empirical relation between outflow current as a function of the rate of loading as depicted in Figure 12.

In all other cases shown in Figures 14b-e, where we re-impacted the tiles within a few minutes, the amount of positive charge flowing out of the stressed subvolume was roughly the same as after the first impact in Figure 14a, but the amount of negative charges flowing back was slightly larger. The difference in the response of a fresh impact versus repeated impacts was probably due to fewer charge carrier recombinations during the short time between drops, only a few minutes. The relatively long "tail" with different slopes in Figure 15 suggests that multiple species of charge carriers were recombining on different time scales.

## 4. Discussion

Stress gradients are created by the tectonic forces that wax and wane in the Earth's crust. During the days, weeks, even months prior to major earthquakes the stresses increase, eventually reaching a point of catastrophic rupture. Recognizing that electric charge carriers are activated in the stressed rock volume and that they can flow out, spreading fast and far, provides insight into the nature of pre-earthquake signals.

Figures 4a/b show an example of a material, Ca-Na-aluminosilicate glass as used in common window panes, which does not provide evidence for any stress-activated charge carriers that could flow out of the stressed central part of a glass plate, producing a measurable current. The 1 mV potential difference that shows up in the moment the stress was applied in this particular experiment was not sufficiently reproducible to warrant further attention. By contrast, whenever we loaded a gabbro tile, reproducible patterns of stress-stimulated outflow current and stress-stimulated surface potentials were observed. The surface potentials and outflow currents were always positive, supporting the conclusion reached in previous publications that the responsible mobile charge carriers are positive.

Positive holes have been reported to be able to flow out of a rock volume, in which they were activated by stress, spreading along the stress gradient into and through adjacent unstressed rock, traveling fast and far [*Freund*, 2002; *Freund et al.*, 2006].

Most published reports of pressure-stimulated currents are based on experiments conducted with rock samples subjected to quasi-uniform loading over their entire cross section (Figure 1b) [*Aydin et al.*, 2009; *Johnston*, 1997; *Kyriazis et al.*, 2009; *Triantis et al.*, 2006; *Tullis*, 2002; *Vallianatos and Triantis*, 2008]. By contrast, in all cases described here we purposefully created stress gradients between the stressed center of our rock tiles and their unstressed edges. We noted that, as soon as we begin loading the center, after a brief negative current, mobile positive charge carriers appear, which are capable of flowing out of the stress subvolume through the unstressed rock to the edges. These charge carriers cause (i) a potential difference between the stressed center and the unstressed edges, i.e. a pressure-stimulated voltage, and (ii) a current the stressed and unstressed volumes, i.e. a pressure-stimulated current. Minor short negative signals were recorded at the beginning and the end of many loading-unloading cycles (Figures 7a/b), but the sign of the dominant signals is persistently positive, indicating that the majority of stress-activated charge carriers in the gabbro tiles are positive. We have focused here on these positive charge carriers.

In semiconductors negative and positive electronic charge carriers have been known for decades. The negative charges are electrons. The positive charges are sites where an electron is missing, hence a defect electron, also known as a "hole" [*del Alamo*, 2011; *Sze*, 1981].

Subjecting the rock to uniaxial stresses leads to the activation of hole charge carriers, which must have existed in the rock prior to the application of stress, albeit in an electrically inactive, dormant state. This is fundamentally different from normal semiconductors where electrons and holes, introduced by doping the materials with aliovalent impurities, are always active. Excess electrons are introduced by N substituting for Si in a silicon crystal, making the silicon n-type. Excess holes are introduced by B substituting for Si to make the silicon p-type *[Sze*, 1981]. In a rock like the gabbro under study here the precursors to the stress-activated holes must have been introduced not by doping but via a fundamentally different route.

Another noteworthy observation during our experiments with the gabbro tiles is that the activation of holes, hence, the currents flowing out of a given rock volume, are not related in a simple way to the stress applied. Low stress levels, <10 MPa, are enough to create surprisingly high outflow currents, indicating that a large number of holes is being activated. However, holes also continue to be activated as upon further loading. When the stresses are constant, the currents decrease slowly, but intermittent additional loading causes transient increases in the outflow currents [*Takeuchi et al.,* 2006]. When the stresses are removed, the outflow currents eventually decrease rapidly, indicating that most of the holes return to an inactive, dormant state. Some take longer. After the samples were annealed for various lengths of time, deactivating the holes, they become reactivated when stresses are reapplied. This procedure can be repeated many times.

Yet another remarkable feature is that, once activated, the holes have the ability to flow out of the stressed subvolume in the center to the unstressed edges of the tiles. The faster the stresses are applied, the larger the outflow currents. When a constant load is maintained, i.e. when a steady stress gradient is maintained between center and the edges of the tile, the hole outflow continues, lasting for hours, even days, weeks and months, indicating very long lifetimes. The outflow will eventually stop when all charge carriers have recombined.

To address these points we need to understand

(i)     where do the charge carriers come from?
(ii)    what is the nature of the stress-activated holes?
(iii)   why are they activated already at very low stress level?
(iv)    why are they able to flow through unstressed rock?
(v)     how long to they live before recombining?

## 4.1 Peroxy Defects and Positive Holes

Among the many different types of point defects in rock-forming minerals, peroxy defects have not received much attention in the geoscience community, in fact practially no attention at all.

As discussed in greater detail in Part I, peroxy defects consist of pairs of covalently bonded oxygen anions in the valence state 1– instead of the usual 2–. First identified in MgO single crystals [*Freund and Wengeler*, 1982] and characterized through a wide range of physical techniques [*Freund et al.*, 1993], evidence for peroxy defects was also found in silica [*Freund and Masuda*, 1991] and silicates [*Freund*, 1985]. A typical peroxy defect in a silicate mineral would be $O_3Si-OO-SiO_3$ replacing the common $O_3Si-O-SiO_3$. Because the $O^--O^-$ distance in the peroxy bond is very short, less than 0.15 nm as compared to typical $O^{2-}-O^{2-}$ distances on the order of 0.28–0.3 nm, the volume occupied by a two $O^-$ in the peroxy bond is comparable to the volume occupied by a single $O^{2-}$ or a single $OH^-$. This implies that high pressures should favor the right hand side of eq. [1] and tend to stabilize peroxy.

As long as peroxy bonds are intact and the two $O^-$ tightly coupled, peroxy defects are electrically inactive. Their presence has little or no effect on the physical properties, specifically on the electrical properties, of the minerals and rocks in which they occur. The small partial molar volume and lack of electrical activity explains why peroxy defects are so inconspicuous and have been consistently overlooked in the past. However, when peroxy bonds break and become activated, the situation changes.

Figure 16 illustrates the way a peroxy defect breaks up in a silicate matrix. The activation of the peroxy bond leads first to a transient state, probably short-lived, where the $O^-–O^-$ bond is decoupled but not yet dissociated [*Freund et al.*, 1993]. The two $O^-$ states are each represented by a dot •, representing a hole[‡]. Next, a neighboring $O^{2-}$ transfers an electron into the decoupled, half-broken peroxy bond. This process consumes one of the holes. As the electron becomes trapped, the $O^{2-}$, which acted as the electron donor, turns into an $O^-$, e.g. into a hole $h^•$.

Figure 16 thus depicts the formation of an electron-hole pair, where the electron e' is trapped in the broken $O^-–O^-$ bond, while the electronic state associated with $O^-$ in a matrix of $O^{2-}$ represents a defect electron or hole, symbolized by $h^•$ [*Freund*, 2011]. Holes associated with monovalent oxygen are different from holes associated with aliovalent substitutional impurities in classical semiconductors, for example boron in silicon, Therefore the hole states associated

---

[‡] In the chemistry literature the dot • is used to designate an unpaired electron. In the Kröger nomenclature for point defects in solids [Kröger, F.A. (1964) *The Chemistry of Imperfect Crystals*, North-Holland, Amsterdam] the dot • is used to designate a hole. Both designations are equivalent because, according to the Pauli Principle, energy levels can be maximally occupied by two electrons. If one is removed, creating a "hole", the one left behind is an unpaired electron.

with O⁻ have been given the name "positive holes" [§] [*Griscom*, 1990].

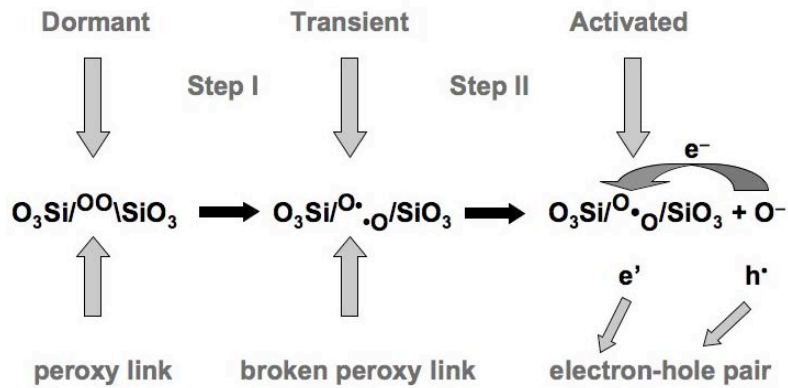

*Figure 16- Schematic representation of the break-up of a peroxy bond in silicate matrix. Step I marks the break-up; Step II marks the electron transfer from an $O^{2-}$ into the decoupled peroxy bond and the generation of an electron-hole pair.*

Figure 17 shows schematically the upper edge of the valence band, which consists primarily of energy levels that derive from O 2sp-type orbitals. Using the designation from Molecular Orbital Theory [*Marfunin*, 1979; *Tossell*, 1983] for the electronic structure of $O_2^{2-}$, the highest occupied levels in the peroxy defect are non-bonding $1\pi_g^{nb}$ levels, while the antibonding $3\sigma_u^*$ level is empty. Thus, the peroxy defect causes a dip in the energy surface at the uppermost edge of the valence band, which is indicated by the solid line.

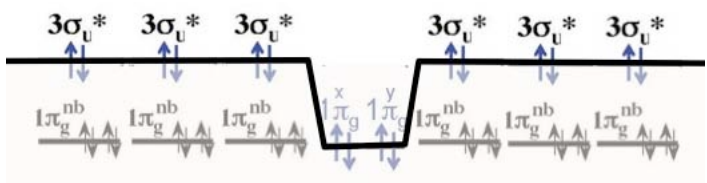

*Figure 17- The energy levels forming the upper edge of the valence band are of O 2sp symmetry. In the case of $O^{2-}$ the highest occupied level is antibonding $3\sigma_u^*$, while in the case of $O_2^{2-}$ it is the non-bonding $1\pi_g^{nb}$ level.*

Figure 18a shows schematically the corresponding energy diagram where the valence and conduction bands are separated by a wide band gap $E_{gap}$. In silicate minerals $E_{gap}$ is several eV – too wide for electrons in the valence band to be thermally activated to the conduction band at the low to moderate temperatures encountered in most of the Earth's crust. Hence, silicate minerals and the rocks, which they form, will not exhibit conductivity by electrons promoted to the conduction band. If rocks exhibit stress-activated electronic conductivity, the underlying process must be fundamentally different. This is underlined by the observation that the stress-activated

---

[§] In semiconductor parlance "holes" are positive charge carriers. The name "positive hole" is redundant.

conductivity of the gabbro tiles studied here is not carried by electrons but by positive holes.

To explain what seems to be happening at the edge of the valence band we need to take a look at the energy levels involved. As indicated in Figure 16 the highest occupied energy levels in the two oxygens making a peroxy bond are of $\pi_g$ symmetry. They derive from the $O2p_x$ and $O2p_y$ orbitals, which form a torus around the $O^-\!-\!O^-$ bond [*Marfunin*, 1979; *Tossell*, 1983], but are non-bonding (superscript "nb"). As long as peroxy defects are intact, their electrons in the $1\pi_g^{nb}$ level are spin-paired and, hence, strictly localized as indicated in Figures 17 and 18a. Break-up of the peroxy bond can be achieved by bending it along the $O^-\!-\!O^-$ axis, thereby lifting the 4-fold degeneracy of the $1\pi_g^{nb}$ levels and causing two of the non-bonding $\pi^{nb}$ electrons to transition into the antibonding $\sigma_u^*$ level. This splits the $1\pi_g^{nb}$ energy level as depicted in Figure 18b, trapping the electron transferred from a neighboring $O^{2-}$ on a new energy level below the edge of the valence band. The remaining hole state forms the delocalized positive hole by mixing the $\pi_g^{nb}$ level with the antibonding $3\sigma_u^*$ level of the surrounding $O^{2-}$ [*Freund et al.*, 1993].

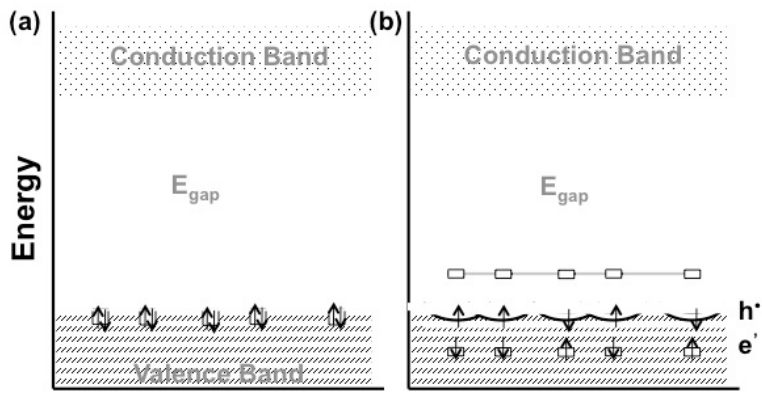

*Figure 18a-  Schematic of the valence and conduction bands with peroxy defects indicated by dips in the upper edge of the VB; Figure 18b- Break-up of the peroxy defects creates new energy levels, –δE below the edge (occupied) and +δE above the edge of the valence band (empty).*

Figure 17b translates the positive hole activation into an energy diagram, where the newly created energy level –δE below the valence band edge, which is occupied by the trapped electron, has a mirror level +δE above the valence band edge that is intrinsically empty as indicated by the open rectangles in Figure 18b. Delocalization of the positive hole is indicated by the cusps at the very edge of the valence band.

At the same time we note that the +δE level above the edge of the valence band is available to be thermally populated by the electrons trapped in the -δE level below the edge of the valence band. The number of electrons on the +δE level is given by Boltzmann statistics $\prod = e^{-[2\delta E/kT]}$, where k is the Boltzmann constant and T the absolute temperature. However, the +δE level above the edge of the valence band is available only within the stressed rock volume, where peroxy defects are broken, creating this pair of energy levels, ±δE level above and below the edge of the valence band. This provides a qualitative understanding as to why the boundary between the stressed and unstressed subvolumes acts as a barrier for electrons:  while positive holes are able to pass through this boundary and propagate through the adjacent unstressed rocks, the electron can not, because there are no ±δE levels available in the unstressed rock.

Since the valence bands of mineral grains in rocks are electronically connected via their O 2sp-type energy levels at the upper edge of the valence band, once positive hole charge carriers are formed, they will be able to propagate through grains and across grain boundaries, subject to only minor scattering, flowing outward well beyond the boundaries of the stressed subvolume. The propagation of positive holes has been proposed to occur by way of a phonon-assisted electron hopping mechanism [*Shluger et al.*, 1992] as depicted in Figure 19. In order for a positive hole to travel from left to right, a succession of electrons has to hop from right to left.

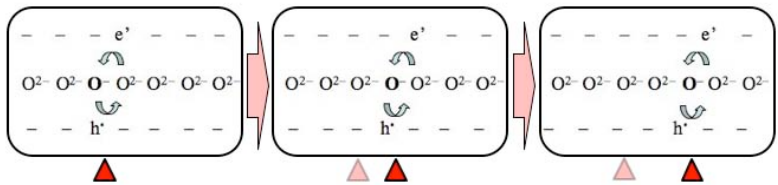

*Figure 19*- *Schematic representation of a positive hole, h•, moving from left to right through a succession of electron hops from right to left.*

Electrons had the highest probability to execute a hop is when the distances between neighboring oxygen anions are shortest. The highest velocity the h• can achieve is thus given by the phonon frequency ν, which modulates the interatomic distances, multiplied with the hopping distance d and divided by a factor of 3 to account for the probability of executing a hop in a given direction. Taking n as $10^{12}$ Hz and the hopping distance d between $O^{2-}$ sites as 0.28 nm, the speed for positive holes to propagate via a phonon-assisted electron hopping mechanism, νd/3, should be on the order of 100 m/sec [*Shluger et al.*, 1992].

In our free-fall impact experiments the peak of the positive outflow current is well resolved, allowing us to estimate the travel time of positive holes to the edges of the tiles. Since we used square tiles, the charge carriers will not flow with equal probability along all possible directions but preferentially along the paths of least resistance.

With a Cu contact along the edge of the tiles, the path of least resistance is the shortest distance from a point in the 5 cm diameter central source area to the closest point on the edge.

Integrating over the circular source area at the center and normalizing over all possible paths, the average path of least resistance is $\frac{L}{2} - \frac{4r\sqrt{2}}{3\pi}$ with r the radius of the central contact, 25 mm, and L the edge length of the tile, 30 cm. The mean path length of least resistance is thus ~13.7 cm.

On impacting the piston in the center of a tile the median time between impact and the maximum of the positive peak was found to be 1.45 ms, corresponding to an h$^\bullet$ velocity of 95±25 m/s, which is fully consistent with the theoretical value.

**4.2 Applying Laboratory Observations to the Field**

When tectonic forces act on rocks deep in the Earth crust subjecting them to increasing levels of stress, e' and h$^\bullet$ charge carriers will be generated through the stress-activation of peroxy defects. As demonstrated in this paper, the h$^\bullet$ charge carriers have the ability to flow out of the rocks subjected to stress. They spread into and through unstressed rocks. By contrast, the co-activated e' charge carriers, i.e. electrons, are mobile only inside the stressed rock volume.

The outflow of h$^\bullet$ charge carriers from a stressed subvolume is driven by the h$^\bullet$ concentration gradient along the stress gradient. The mechanism by which the h$^\bullet$ propagate does not depend on the presence of intergranular carbon films nor on the presence of intergranular connected fluids. However, to the extent that intergranular carbon films are present, they may play a role in closing the circuit by allowing electrons to also flow out. Their outflow via carbon films would the equivalent to the outflow of electrons in our laboratory experiments depicted in Figures 1c and 2b via the Cu wire connecting Cu contacts attached to the stressed subvolume and to the unstressed end. If, by contrast, the rocks contain networks of interconnected intergranular aqueous films, circuit closure could be achieved by hydronium ions, $H_3O^+$, or cations flowing into the stressed subvolume from the intergranular space.

The electrical conductivity of igneous rocks increases exponentially with temperature over the 300-650°C window, characterized by an activation energy around 1 eV [*Parkhomenko and Bondarenko*, 1986], suggestive of h$^\bullet$ as dominant charge carriers. Thus, in the field, along the geotherm, the outflow of stress-activated h$^\bullet$ can be expected to increase with depth. This should allow for higher h$^\bullet$ current densities flowing out of stressed rock volumes deeper in the crust.

Any flow of h$^•$ through the rocks constitutes an electric current. If circuit closure can be achieved as discussed here, the h$^•$ outflow along stress gradients may form a current that can persist for a period of time, either in a quasi-dc mode or fluctuating. A dc outflow current that waxes and wanes slowly, over a long time period, would lead to a drift of the regional magnetic field strength, possibly measurable at the Earth surface.

For instance, before the magnitude 7.6 Chi-Chi earthquake of September 21, 1999, an 8-station network of total magnetic field sensors in operation in Taiwan since 1989 [*Chen et al.*, 2004] recorded a slow, but steady drift of the magnetic field over the Northwestern part of the island, starting in 1997. It resulted in an "anomaly" relative to the International Geomagnetic Reference Field [*Finlay and S. Maus*, 2010], the maximum of which reached about 250 nT and was concentrated where the Chelungpu Fault would rupture during the Chi-Chi event. Seven weeks before the Chi-Chi earthquake the magnetometer station closest to the Chelungpu Fault started to record very strong magnetic field fluctuations with amplitudes up to 250 nT, suggesting that the powerful current below had transitioned from a quasi-dc mode to an ac mode, leading to ultralow frequency (ULF) emissions in the range of milliHertz [*Freund and Pilorz*, 2012; *Yen et al.*, 2004]. Increased ULF activity has been recorded at the Earth surface before other major earthquakes as well [*Bleier et al.,* 2009; *Chavez and Millan-Almaraz,* 2010; *Hall et al.,* 2004; *Sharma et al.*, 2011; *Smirnova et al.,* 2013].

One of the findings of our study presented here is that the amount of charge flowing out of a given volume of gabbro during a period of increasing stress appears to be relatively constant, about $10^6$ Coulomb km$^{-3}$ during loading from 0 MPa to about 60 MPa, irrespective of how fast the stresses are applied. If the stresses are applied within 1 sec, the current outflow will be on the order of $10^6$ A km$^{-3}$. In the free-fall experiments, central loading of an 18 cm$^3$ subvolume at the center of the tile within about 1-1.5 ms led to a current outflow on the order of 20-30 μA. Linearly extrapolating this values to a rock volume of 1 km$^3$ would give ~$10^9$ A.

## 5. Conclusions

The experimental work presented here shows that loading the center of gabbro tiles leads to electric currents that flow out of the stressed subvolume to the unstressed edges. A potential difference develops between the center and the edges, which drives the outflow currents already at low stress levels, >1 MPa. There are brief transient negative voltages and currents, but the main outflow currents are consistently positive. They grow with the applied load, rapidly at first, going through maxima and minima, then increase steadily as the load is increased at constant rates to stress levels between 48 MPa to 64 MPa, about ¼ of the fracture strength of the gabbro under study. If the load is kept constant, the currents continue to flow for hours, days, weeks, even months, slowly decreasing with time. Increasing the loading rates over 5 orders of magnitude, from 0.2 kPa/sec to 20 MPa/sec, causes the peak outflow currents to also increase. The integrated charge flowing out was found to be on the order of $10^6$ Coulomb $km^{-3}$, relatively independent of the rate at which the stresses are applied.

The rate of loading was increased by another 3 orders of magnitude by drop tower experiments, stressing the rock tiles parabolically within 0.5-2 ms without fracture. Very large positive outflow currents are recorded, lasting 4-6 ms and reaching up to 30 µA flowing out of less than 20 $cm^3$ of rock. Linearly extrapolating this to 1 $km^3$ of rock gives a value of 1-2 x $10^9$ A $km^{-3}$.

It appears that no other research groups ever conducted experiments, where only a subvolume of rock is subjected to stress. The most likely reason is that the flow of stress-activated electric currents from a stressed rock volume into and through unstressed rocks was deemed impossible. Hence, no attempts were made to measure stress-stimulated currents and voltages with the appropriate sample geometry (Figure 1c/d). Setting up such experiments became sensible only after a basic understanding had been achieved about the nature of peroxy defects and about their ability to generate positive hole charge carriers, which can flow out from stressed rocks into and through unstressed rocks. Though peroxy defects are ubiquitous, their presence and influence – possibly controlling influence – on the electrical properties of rocks have not yet attracted the level of attention in the geoscience community that they probably deserve.

**Acknowledgments**

The early part of this work was supported through a GEST (Goddard Earth Science Technology) Fellowship and later through a grant NNX12AL71G from the NASA Earth Surface and Interior (ESI) program to Friedemann Freund. We thank Dr. Akihiro Takeuchi and Dr. Bobby S Lau for valuable help during the experimental part of this study. We thank Professor Charles Schwartz, University of Maryland, Department of Civil Engineering, for access to the hydraulic press. We thank Hollis H Jones, NASA Goddard Space Flight Center, for help with the analysis of the experimental data. We thank Lynn Hofland, NASA Ames Research Center, Engineering

Evaluation Laboratory (EEL), for his assistance during the drop tower experiments. We thank AJ Udom for his participation as part of his NASA Ames Summer 2013 Internship, and we thank Gary Cyr for help with setting up the data acquisition system.**References**

Anderson, G., and C. Ji (2003), Static stress transfer during the 2002 Nenana Mountain-Denali fault, Alaska, earthquake sequence, *Geophysical Research Letters*, *30*(6), 1310.

Aydin, A., R. J. Prance, H. Prance, and C. J. Harland (2009), Observation of pressure stimulated voltages in rocks using an electric potential sensor, *Appl. Phys. Lett.*, *95*, 124102.

Bleier, T., C. Dunson, M. Maniscalco, N. Bryant, R. Bambery, and F. T. Freund (2009), Investigation of ULF magnetic pulsations, air conductivity changes, and infra red signatures associated with the 30 October 2007 Alum Rock M5.4 earthquake, *Nat. Hazards Earth Syst. Sci.*, *9*, 585-603.

Berlin, H. M., and F. C. Getz (1988), *Principles of Electronic Instrumentation and Measurement*, Merrill Pub. Co.

Chavez, O., and J. R. Millan-Almaraz, Pérez-Enríquez, R., Arzate-Flores, J. A., Kotsarenko, A., Cruz-Abeyro, J. A., and Rojas, E. (2010), Detection of ULF geomagnetic signals associated with seismic events in Central Mexico using Discrete Wavelet Transform, *Nat. Hazards Earth Syst. Sci.*, *10*, 2557-2564.

Chen, C. H., J. Y. Liu, H. Y. Yen, X. Zeng, and Y. H. Yeh (2004), Changes of geomagnetic total field and occurrences of earthquakes in Taiwan, *Terr. Atmo. Ocean. Sci.*, *15*, 361-370.

Finlay, C. C., and C. D. B. S. Maus, T. N. Bondar, A. Chambodut, T. A. Chernova, A. Chulliat, V. P. Golovkov, B. Hamilton, M. Hamoudi, R. Holme, G. Hulot, W. Kuang, B. Langlais, V. Lesur, F. J. Lowes, H. Luhr, S. Macmillan, M. Mandea, S. McLean, C. Manoj, M. Menvielle, I. Michaelis, N. Olsen, J. Rauberg, M. Rother, T. J. Sabaka, A. Tangborn, L. Toffner-Clausen, E. Thebault, A. W. P. Thomson, I. Wardinski Z. Wei, T. I. Zvereva. (2010), International Geomagnetic Reference Field: the eleventh generation, *Geophys. J. Int.*, *183*(3), 1216-1230.30

Freund, F., and H. Wengeler (1982), The infrared spectrum of OH--compensated defect sites in C-doped MgO and CaO single crystals, *J. Phys. Chem. Solids*, *43*, 129-145.

Freund, F. (1985), Conversion of dissolved "water" into molecular hydrogen and peroxy linkages, *J. Non-Cryst. Solids*, *71*, 195-202.

Freund, F., and M. M. Masuda (1991), Highly mobile oxygen hole-type charge carriers in fused silica, *J. Mater. Res.*, *6*(8), 1619-1622.

Freund, F., M. M. Freund, and F. Batllo (1993), Critical review of electrical conductivity measurements and charge distribution analysis of magnesium oxide, *J. Geophys. Res.*, *98*(B12), 22209-22229.

Freund, F. (2002), Charge generation and propagation in rocks, *J. Geodynamics*, *33*, 545-572.

Freund, F., and S. Pilorz (2012), Electric Currents in the Earth Crust and the Generation of Pre-Earthquake ULF Signals, in *Frontier of Earthquake Prediction Studies*, edited by M. Hayakawa, pp. 464-508, Nippon Shuppan, Tokyo.

Freund, F. T., A. Takeuchi, and B. W. Lau (2006), Electric currents streaming out of stressed igneous rocks - A step towards understanding pre-earthquake low frequency EM emissions, *Physics and Chemistry of the Earth*, *31*(4-9), 389-396.

Freund, F. T., and D. Sornette (2007), Electromagnetic earthquake bursts and critical rupture of peroxy bond networks in rocks, *Tectonophys.*, *431*, 33-47.

Freund, F. T. (2011), Pre-Earthquake Signals: Underlying Physical Processes, *Journal of Asian Earth Sciences*, *41*, 383–400.

Fuji-ta, K., T. Katsura, and Y. Tainosho (2004), Electrical conductivity measurement of granulite under mid- to lower crustal pressure-temperature conditions, *Geophys. J. Inter.*, *157*, 79-86.

Glover, P. W. J. (1996), Graphite and electrical conductivity in the lower continental crust: A review, *Physics and Chemistry of The Earth*, *21*(4), 279-287.

Gorshkov, M. M., V. T. Zaikin, and S. V. Lobachev (2001), Electrical Conductivity of Rocks under Shock Compression, *Journal of Applied Mechanics and Technical Physics*, *42*, 196-201.

Green, D. H., and A. E. Ringwood (1967), The genesis of basaltic magmas, *Contributions to Mineralogy and Petrology*, *05*(2), 103-190.

Griscom, D. L. (1990), Electron spin resonance, *Glass Sci. Technol.*, *48*, 151-251.

Hall, C. G., H. Y. Yen, H. C. Chen, A. Takeuchi, B. W. Lau, and F. Freund (2004), Anomalous Magnetic Field Pulses, Ground Currents, and the Build-up of Stress prior to the Chi-Chi Earthquake, American Geophysical Union, Fall Mtg. 2004, San Francisco.

Heikamp, S., and G. Nover (2001), The electrical signature of rock samples exposed to hydrostatic and triaxial pressures, *Annalie di Geofisica*, *44*(2), 287-293.

Johnston, M. J. S. (1997), Review of electric and magnetic fields accompanying seismic and volcanic activity, *Surveys in Geophysics*, *18*, 441-475.

Kyriazis, P., C. Anastasiadis, I. Stavrakas, D. Triantis, and J. Stonham (2009), Modelling of electric signals stimulated by bending of rock beams, *International Journal of Microstructure and Materials Properties*, *4*(1), 5-18.

Marfunin, A. S. (1979), *Spectroscopy, Luminescence and Radiation Centers in Minerals*, 257-262 pp., Springer Verlag, New York.

Nover, G. (2005), Electrical Properties of Crustal and Mantle Rocks – A Review of Laboratory Measurements and their Explanation, *Surveys in Geophysics*, *26*(5), 593-651.

Nover, G., J. B. Stoll, and J. von der Gönna (2005), Promotion of graphite formation by tectonic stress – a laboratory experiment, *Geophysical Journal International*, *160*(3), 1059-1067.

Parkhomenko, E. I., and A. T. Bondarenko (1986), *Electrical conductivity of rocks at high pressures and temperatures*, Elektroprovodnost Gornykh Porod Privysokikh Davleniyakh i Temperaturakh, Moscow 1972 ed., 212 pp., NASA.

Sharma, A. K., P. A. V., and R. N. Haridas (2011), Investigation of ULF magnetic anomaly before moderate earthquakes, *Exploration Geophysics 43*(1), 36-46.

Smirnova, N. A., D. A. Kiyashchenko, V. N. Troyan, and M. Hayakawa (2013), Multifractal approach to study the earthquake precursory signatures using the ground-based observations,,

*Review of Applied Physics*, *2*(3), 58-67.

Shluger, A. L., E. N. Heifets, J. D. Gale, and C. R. A. Catlow (1992), Theoretical simulation of localized holes in MgO, *J. Phys.: Condens. Matter*, *4*(26), 5711-5722.

Takeuchi, A., B. W. Lau, and F. T. Freund (2006), Current and surface potential induced by stress-activated positive holes in igneous rocks, *Physics and Chemistry of the Earth*, *31*(4-9), 240-247.

Tossell, J. A. (1983), A qualitative molecular orbital study of the stability of polyanions in mineral structures, *Phys. Chem Minerals*, *9*(3-4), 115-123.

Triantis, D., and I. S. Filippos Vallianatos, George Hloupis (2012), Relaxation phenomena of electrical signal emissions from rock following application of abrupt mechanical stress, *Annals of Geophysics*, *55*(1), 207-212.

Triantis, D., I. Stavrakas, C. Anastasiadis, A. Kyriazopoulos, and F. Vallianatos (2006), An analysis of pressure stimulated currents (PSC), in marble samples under mechanical stress, *Physics and Chemistry of the Earth*, *31*(4-9), 234-239.

Tullis, J. (2002), Deformation of Granitic Rocks: Experimental Studies and Natural Examples, in *Plastic Deformation of Minerals and Rocks*, edited by S.-I. Karato and H.-R. Wenk, pp. 51-95, Mineralogical Society of America.

Vallianatos, F., and D. Triantis (2008), Scaling in Pressure Stimulated Currents related with rock fracture, *Physica A: Statistical Mechanics and its Applications*, *387*(19-20), 4940-4946.

Yen, H.-Y., C.-H. Chen, Y.-H. Yeh, J.-Y. Liu, C.-R. Lin, and Y.-B. Tsai (2004), Geomagnetic fluctuations during the 1999 Chi-Chi earthquake in Taiwan, *Earth Planets Space*, *56*, 39-45.

Zoback, M. D., M. L. Zoback, V. S. Mount, J. Suppe, J. P. Eaton, J. H. Healy, D. Oppenheimer, P. Reasenberg, L. Jones, C. B. Raleigh, I. G. Wong, O. Scotti, and C. Wentworth (1987), New Evidence on the State of Stress of the San Andreas Fault System, *Science*, *238*, 1105-1111.